\documentclass{article}

\usepackage{arxiv}

\usepackage[utf8]{inputenc} 
\usepackage[T1]{fontenc}    
\usepackage{hyperref}       
\usepackage{url}            
\usepackage{booktabs}       
\usepackage{amsfonts}       
\usepackage{nicefrac}       
\usepackage{microtype}      
\usepackage{lipsum}
\usepackage{graphicx}
\usepackage{listings}
\graphicspath{ {./images/} }

\title{Retrieval-Augmented Generation to Support Railways Engineering Tasks: A Case Study}

\author{
  Andrea Gerardo Russo \\
  NIER Engineering S.p.A.\\
  \texttt{a.russo@nier.it} \\
  \And
  Federico Ruggeri \\
  University of Bologna \\
  \texttt{federico.ruggeri6@unibo.it}\\
  \And
  Ivan Tomarchio \\
  NIER Engineering S.p.A.\\
  \texttt{i.tomarchio@nier.it}\\
  \And
  Davide Bombini \\
  NIER Engineering S.p.A.\\
  \texttt{d.bombini@nier.it}\\
  \And
  Nicolò Donati \\
  University of Bologna \\
  \texttt{n.donati@unibo.it} \\
  \And
  Gianmarco Pappacoda \\
  University of Bologna \\
  \texttt{gianmarco.pappacoda@unibo.it} \\
  \And
  Paolo Torroni \\
  University of Bologna \\
  \texttt{p.torroni@unibo.it} \\
  \And
  Giuseppe-Emiliano La Cara \\
  NIER Engineering S.p.A.\\
  \texttt{e.lacara@nier.it}
}

\begin{document}
\maketitle
\begin{abstract}
The growing number and complexity of technical regulations represent an important challenge for all professionals in regulated industries. This paper describes a case study, from design to deployment, of building a Retrieval-Augmented Generation system for the consultation of complex technical regulations in the railway domain. Although developed for the railway sector, this testimony of an industrial experience is of particular value for technical domains where regulatory compliance and accurate information retrieval from complex documentation are essential requirements. It also constitutes a human-centered approach for implementing LLM-powered technical documentation consultation across various regulated industries, balancing technological capabilities with domain expertise.  
\end{abstract}


\section{Introduction}
Companies and industries selling their goods in the European market must often comply with a set of technical standards and regulations aimed at guaranteeing European citizens safe and high-quality products. 
Over the years, the European Union has issued regulations regarding several aspects in many industrial sectors, such as biomedical  \cite{europeanbiomedicalreg}, chemical \cite{eu-chemical-reg}, automotive \cite{eu-car-reg} and railway \cite{eu-railway-reg}. 
These technical regulations address all aspects of the product life cycle, from design to decommission, in a very detailed manner.
They are typically written in highly specialized language and contain lengthy, detailed tables and complex diagrams. 
Therefore, their consultation and correct interpretation can be time-consuming even for experienced professionals. 

The advent of generative unlocked new possibilities. 
In recent years,  LLMs have reported outstanding
performance in general question-answering tasks~\cite{petroni-etal-2019-language} and their results are even better when using in-context learning with task instructions and few-shot demonstrations \cite{long2023adaptcontextsretrievalaugmenteddomain}. 
However, as they are bound to a fixed knowledge base that cannot be easily updated \cite{wu2024continuallearninglargelanguage}, they cannot provide responses about topics not contained in their training dataset. 
More importantly, due to their probabilistic nature, it has been observed that LLM-generated responses can be factually incorrect and even nonsensical~\cite{xu2025hallucinationinevitableinnatelimitation}, causing skepticism in business and industrial domains where factual consistency is key~\cite{xu2025hallucinationinevitableinnatelimitation}. 
An interesting approach to mitigating these issue is Retrieval-Augmented Generation (RAG) \cite{lewis2021retrievalaugmentedgenerationknowledgeintensivenlp}, which extends generative models with a retrieval component, whereby document chunks relevant to a given user query are retrieved from an external knowledge base and inserted into an augmented prompt \cite{gao2024retrievalaugmentedgenerationlargelanguage}. 
RAG has been successfully applied in many technical domains such as finance \cite{liu2023fingptdemocratizinginternetscaledata} and medicine \cite{xiong2024improvingretrievalaugmentedgenerationmedicine}. 

This work describes a case study, from design to deployment, of a RAG implementation for the railway sector. The end-users of this framework are engineering professionals whose job is to develop systems that comply with the technical standards and specifications that railway products sold in the European Union are required to comply with, to ensure maximum safety. For example, the railways interoperability defined in the context of the ``European Rail Traffic Management System'' (ERTMS) \cite{ertmspaper} is regulated by the technical specifications issued by the UNISIG consortium. These specifications present complex features such as extensive tables, detailed diagrams, cross-referenced sections, and highly specialized technical language. These are unique challenges for LLM applications as their misinterpretations could compromise system safety. In fact, the reliability of the responses is a critical feature requested by engineering professionals with different levels of experience (they span from juniors with limited experience to seniors with decades of knowledge) and constantly consult these specifications across the several steps of the product life-cycle. 
It is a challenging case study due to its complexity and the required high levels of safety by both national and international regulatory entities. 

The project we describe is the result of a joint effort between company employees and academic researchers. 
A six-phase approach was implemented to provide a useful and reliable tool for an engineering company in the railway sector and to create a robust and easy-to-follow workflow for future implementation in other business and industrial contexts.
Our case study can be used as a blueprint by engineering companies that need to design solutions for the consultation of technical specifications and standards by using generative AI techniques.

\section{Development Methodology}
Our methodology is articulated into six phases: requirements definition, prototyping, LLM selection, data collection, fine-tuning, testing/validation. 
The following sections provide a detailed description of each phase, including specific methodologies, challenges encountered, and solutions developed throughout the process. 

\subsection{Requirements definitions}
\label{phase1}
Part of the requirements definition was the selection of relevant technical specifications to become the system's knowledge based.
We selected three UNISIG technical specifications, in the form of PDF documents in English: SUBSET-026, SUBSET-037 and SUBSET-098. 
This limited set well illustrates the complexity that strongly characterizes the railway sector and its underlying regulations. 
Indeed, they are very long documents (e.g., SUBSET-026 is 701 pages long and is divided into nine different chapters), they contain tables that extend across multiple pages, diagrams in the form of images and highly specialized terminology. 
Moreover, the end-users expressed the need to have a tool capable of answering also complex questions that typically require combining information from separate and distant paragraphs in the specifications. Finally, due to the ubiquitous presence of non-disclosure agreements (NDA) in the business and industrial context, the end-users requested full control of the data flow. 

\subsection{Prototyping}
\label{phase2}
To address the requirements defined in Section~\ref{phase1}, the implementation of a RAG framework powered by an open-weights LLM was planned. 
To avoid additional translation steps that could introduce potential risks \cite{wang-etal-2024-languages}, end-users opted to interact with the framework in the language of the documentation, English, and not in their native language. 
A on-premise installation was selected with respect the use of online services to meet the requested level of data protection and privacy.
To further reduce the energy footprint and the costs associated with the management of this type of installation, the selection of the right LLM was constrained to small and mid-size models, up to 7B.
Finally, a fine-tuning process with hand-annotated data was planned to deal with the specific terminology, the long complex tables, and the style of interaction expected by the end-users.   

The RAG framework was implemented in Python using the \textit{LangChain} library.\footnote{\url{https://python.langchain.com}} The presence of tables and figures was addressed by using the python library \textit{Unstructured}.\footnote{\url{https://github.com/Unstructured-IO/unstructured}}.
A FAISS database~\cite{douze2025faisslibrary} was used as vector store and retriever. 
In agreement with the end-users, only plain-text and tables were kept while images were discarded. The rationale behind this choice is that, at the beginning of the project (early 2024), the abilities of multi-modal models were still not satisfying for a business context, especially for little and mid-size LLMs. 
During the testing of the early version of the framework, end-users expressed the necessity to go beyond single question-answer interaction and to have a protection against out-of-domain questions and other improper use. 
Therefore, the first need was addressed by powering the LLM with a short-memory mechanism, whereas the second was fulfilled introducing a guardrail (see Appendix~\ref{app:guardrail} for more details).

\subsection{LLM selection}
\label{phase3}
Given the end-users' need to host the framework on a local machine, the only possible candidate were open-weights models with a commercial use license. 
To select the LLM to be used in the framework, a group of 10 end-users, with extensive knowledge of UNISIG specifications and representing a good sample of the company's workforce, was involved in an evaluation process.
First, this group (called Evaluation Group) was requested to write at least 5 questions regarding the three UNISIG specifications used in this work, thus resulting in a dataset of 67 question-answer pairs. 
To further facilitate the evaluation, the evaluation group was also requested to annotate, for each question-answer pair, the words of the associated UNISIG paragraph, thus creating the so called Evaluation Dataset. 
Then, considering the available LLMs at the time of this phase (March-May 2024), the following LLMs were selected from the \textit{HuggingFace} repository: \textit{Zephyr-beta}\footnote{{\texttt{zephyr-7b-beta}}}, \textit{Mistral}\footnote{\texttt{mistralai/Mistral-7B-Instruct-v0.2}} and \textit{Falcon}\footnote{\texttt{tiiuae/falcon-7b}}.

Each LLM was embedded into the implemented RAG framework and provided with the questions (one at a time) contained in the evaluation dataset as input. Finally, the Evaluation Group was requested to evaluate the collected LLM-generated responses. In particular, they were asked to rate the responses on a scale from 0 to 2, where a score of 0 indicated a completely wrong response, a score of 1 indicated a partially correct response, and a score of 2 indicated a completely correct one. To avoid possible biases, no information about which LLM generated each response was given to the annotators. Additionally, annotators were asked to repeat the evaluation process twice, where, in each round, they were assigned randomly selected samples to annotate.

\subsection{Data collection}
\label{phase4}
\begin{figure}[t]
\centering
\includegraphics[width=\linewidth]{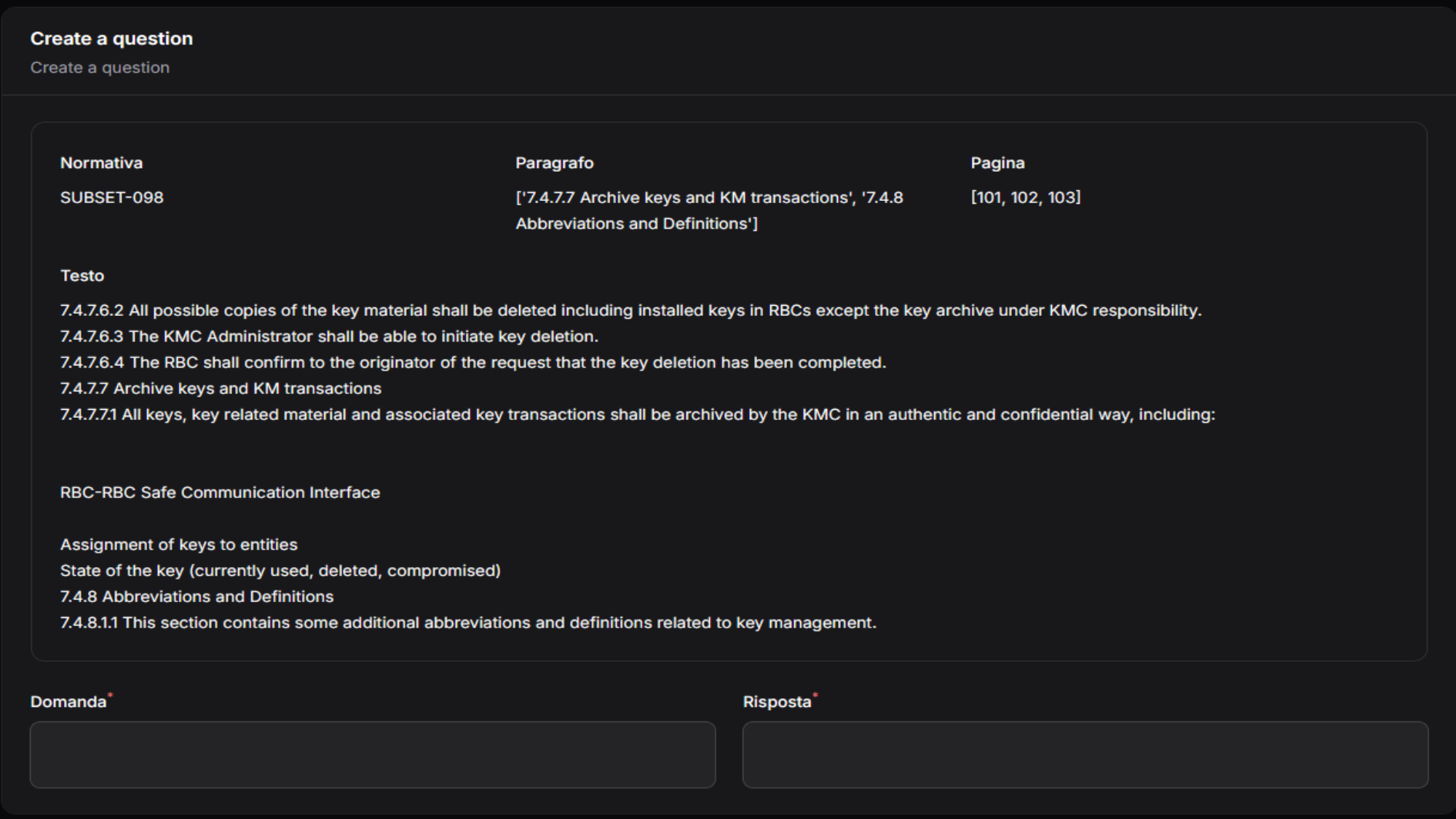}
\caption{Examples of the annotation process using the GeDI tool.}\label{fig:GeDI}
\end{figure}

General-purpose LLMs typically lack deep knowledge in specific fields. 
Therefore, to improve the performance of the LLM selected in Section~\ref{phase3}, 15 end-users (some of whom were already in the Evaluation Group) were involved in the creation of a domain-specific dataset to be used as input in the next phase. 
This group, called the Annotation Group, includes engineering professionals for whom the consultation of the UNISIG specifications is a daily routine.
To facilitate their task and speed up and standardize the whole process, the chunks obtained from the UNISIG specifications (see Section~\ref{phase2}) were used as the internal database of a web-based tool that was specifically implemented for this purpose (see Appendix~\ref{app:extraction} for the technical details of the implementation).
Each member of the Annotation Group was provided with an account and was asked to create at least 40 question-answer pairs. 
In detail, after the logging, each annotator was presented with a chunk randomly selected from the internal database and was requested to insert a question related to the displayed chunk and the corresponding answer (see Figure \ref{fig:GeDI}). 
For each chunk, the number of pages and the name of the specification were displayed to help the annotator retrieve additional context when in doubt. 
At the end of the process, the annotated data were exported in a csv file and carefully evaluated.
Hereafter, this dataset is called the Human Dataset. 

\subsection{Fine-tuning and graphical interface creation}
\label{phase5}
The requirement definition phase highlighted the need for an LLM capable of dealing with domain-specific terminology and a response layout aligned to the style of the end-users. 
Therefore, Zephyr LLM underwent a fine-tuning process divided into two steps and involving the following datasets (see Appendix~\ref{app:datasets} for more details):

\begin{enumerate}
    \item \textbf{Artificial Dataset}: this dataset was created using a third independent LLM, namely \textit{Qwen2.5-14B}. 
    This LLM was asked to generate a question-answer pair for each correctly parsed sentence of UNISIG specifications.
    The obtained dataset underwent a post-processing steps to remove questions where there was an explicit reference to the context (e.g.,``\textit{What is the purpose of the MAC value computed using the CBC-MAC function in the given context?}'') or responses with wrong formatted acronym (e.g., ``\textit{What does Pr\% represent in the context of time accuracy parameters?}'') thus resulting in a dataset containing 323 pairs for UNISIG-098, 342 UNISIG-037 and 2166 UNISIG-026 for a total of 2831 questions-answer pairs.
    
    \item \textbf{Human Dataset}: this dataset is the result of the activities performed in Section~\ref{phase4}. It is composed of 665 question-answer pairs.
    
    \item \textbf{Evaluation Dataset}: this dataset is the result of the activity performed in Section~\ref{phase3}. 
    The dataset is composed of 68 question-answer pairs.
\end{enumerate}

The selected LLM Zephyr underwent two supervised fine-tuning processes. 
The first used the Artificial Dataset as training dataset and aimed at increasing the railway-related knowledge of the LLM. The second used the Human Dataset as input and it was based on the Retrieval Augmented Fine-Tuning (RAFT) process \cite{zhang2024raftadaptinglanguagemodel}, aiming at improving the LLM's ability in selecting only the relevant chunks (among the retrieved ones) to generate the response.
In our case, to implement the RAFT process, the Human Dataset was adapted to have examples composed of three elements: the question, the expected answer, and a context formed by the related correct chunk and a distracting one (i.e., a chunk completely unrelated to the question-answer pair). 
In both steps, the LoRA approach \cite{hu2021loralowrankadaptationlarge} has been used (see Appendices~\ref{app:training} and~\ref{app:quantization} for model details).  

At the end of this whole fine-tuning process, the resulting model underwent a quantization procedure \cite{egashira2024exploitingllmquantization} to reduce the LLM-associated computational costs \cite{li2024evaluatingquantizedlargelanguage} and, thus, facilitate the need of the end-user of having an on-premise installation.
Therefore, the floating precision of the fine-tuned LLM was reduced by a factor of 4 (i.e., from 16 bytes to 4 bytes), as it has been shown that this does not significantly affect the performance \cite{li2024evaluatingquantizedlargelanguage}. 

The effect of both the fine-tuning and the quantization on the performance of the LLM has been evaluated on the Evaluation Dataset by providing the LLM with the correct context (i.e., the one associated with the answer-question pair).
The obtained responses were rated automatically using ROUGE \cite{lin-2004-rouge}, whereas a custom 4-point Likert scale was used in the manual evaluation.
In particular, a domain expert evaluated the generated responses of the LLM in three different versions: non-fine-tuned, fine-tuned, and quantized fine-tuned.
As in Section~\ref{phase3}, no information about which LLM generated each response was given to the evaluators.
Moreover, in case of doubts, the evaluator had the possibility to compare the generated response with the ground truth available in the dataset. 

\begin{figure}[!tb]
\centering
\includegraphics[width=\linewidth]{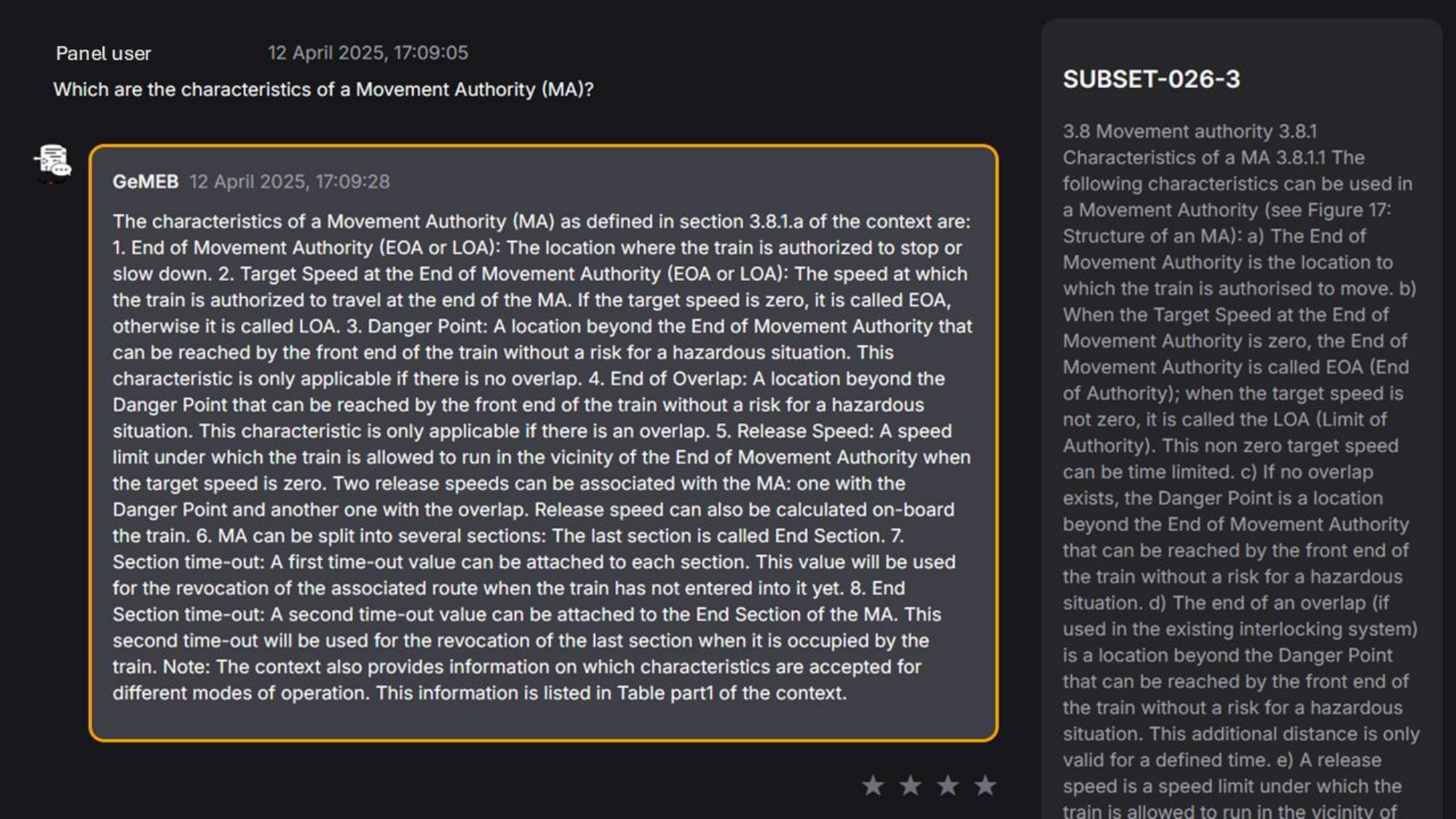}
\caption{Examples of interaction with the graphical interface.}\label{fig:chat_example}
\end{figure}

Besides the fine-tuning to best align the LLM to the task, the end-users requested an easy-to-use interface for the framework. Therefore, we implemented a web-based application accessible only to the local network.
The web-based application provides two authentication levels: panel user and administrator. 
The panel user can interact with the LLM by selecting the UNISIG specification they want to interrogate (i.e., SUBSET-026, SUBSET-037, SUBSET-098, or all of them) and then typing a question in a specific chatbox. 
For each response received, the chunks of the considered UNISIG specification, used to generate the response, are displayed on the right side of the interface to provide to the user more context and facilitate the evaluation. Moreover, the user can rate each response on a scale of 1 to 4 (with higher numbers indicating better response) and motivate the score by typing a comment in the dedicated box (see Figure \ref{fig:chat_example} for an example of chat).
The user with administrator privileges can interact with the LLM in the same way as the panel user, but, in addition, they can also access the chats of all the panel users (previously anonymized) and export them in a csv file for diagnostic purposes. 
More details about the implementation and the usage of the graphical interface are reported in Appendix~\ref{app:gui}.

\subsection{Testing and validation}
\label{phase6}
A group of 5 users was involved in evaluating the model performance in real-world scenarios. This group, called the Testing Group, was a good representation of the typical end-user as they had no experience in generative AI and were completely blind to the design choices and the internal functioning of the framework.
This group was clearly instructed and performed two rounds of evaluation.
In particular, the Testing Group was requested to evaluate both the quality of the received responses and the functionalities of the graphical interface. 
The LLM performance was evaluated by first using the questions contained in the Evaluation Dataset and then by using additional questions not contained in the dataset. The testers were asked to rate each response using the framework's rating system (see Figure~\ref{fig:chat_example}) and, when in doubt, consult the related UNISIG specification. Moreover, when using the questions of the Evaluation Dataset the testers had also the possibility to compare the LLM generated response with the expected one that was previously inserted by the Evaluation Group. On the other hand, the testing of the graphical interface functionalities was not guided by a specific procedure, as the testers were asked to behave as if in a real-world scenario and report anomalous situations. 
In addition, a User Manual of the web-based tool was issued and provided to the Testing Group as reference.

For each evaluation round, the Testing Group released a report containing the observed issues and optionally a set of suggestions for new functionalities that was used to release a modified version of the framework. 
In total three different versions of the framework were released (see Appendices~\ref{app1} and~\ref{app2} for more details).

\section{Results}
\subsection{LLM selection}
In a preliminary analysis, the responses of the \textit{Falcon} model were excluded from the above-described selection process due to their low quality (e.g., presence of long sequences of repeated words).
The remaining two models showed an equal number of score 2 responses, although \textit{Zephyr} yielded better values for both score 0 and score 1 responses  (see Table \ref{table:Table 1}). For this reason, we selected Zephyr-7B-beta as the LLM to use in the project. 

\begin{table}[bt!]
\centering
\begin{tabular}{ c r r r }
\toprule
LLM     & \multicolumn{1}{c}{Score 2} & \multicolumn{1}{c}{Score 1} & \multicolumn{1}{c}{Score 0} \\ 
\midrule
\multicolumn{4}{c}{\textbf{Round 1}}           \\ 
\midrule
Mistral & 19      & 25      & 24      \\  
Zephyr  & 15      & 36      & 17      \\ 
\midrule
\multicolumn{4}{c}{\textbf{Round 2}}           \\  
\midrule
Mistral & 17      & 24      & 27      \\  
Zephyr  & 21      & 32      & 15      \\  
\midrule
\multicolumn{4}{c}{\textbf{Total}}             \\  
\midrule
Mistral & 36      & 49      & 51      \\  
Zephyr  & 36      & 68      & 32 \\
\bottomrule
\end{tabular}
\caption{Results of the LLM selection process.}\label{table:Table 1}
\end{table}

\subsection{Data collection output}
The data collection process lasted approximately one month and resulted in a dataset of 665 entries. We identified 40 question-answer pairs (about 6\% of the total) that were considered ill-formed. For example, in some cases the answer was composed by just one word, whereas in other cases the question-answer pair was written using a mix of English and the annotator's native language. In most cases, quick fixes were implemented, such as translating sentences to English. In other cases, the annotator was contacted to provide explanations and find a proper fix for the issue. 

\subsection{Evaluation of fine-tuning and quantization}
The fine-tuning on the Augmented Dataset reached a minimum loss value of 1.5 on the validation dataset at the third epoch (Table \ref{tab:augmented_loss}), whereas the fine-tuning on the Human Dataset reached the minimum loss value of 1.05 on the validation dataset at the ninth epoch (see Table \ref{tab:human_loss}).

\begin{table}[!tb]
\centering
\begin{tabular}{l r r}
\toprule
\textbf{Epoch} & \multicolumn{1}{c}{\textbf{Train loss}} & \multicolumn{1}{c}{\textbf{Validation loss}} \\ 
\midrule
1              & 2.39                                     & 1.73                                          \\ 
2              & 1.51                                     & 1.57                                          \\ 
3              & 1.01                                     & 1.50                                          \\ 
4              & 0.74                                     & 1.59                                          \\ 
5              & 0.55                                     & 1.74                                          \\ 
\bottomrule
\end{tabular}
\caption{Loss values during the supervise fine-tuning on the Augmented Dataset.}
\label{tab:augmented_loss}
\end{table}

\begin{table}[tb]
\centering
\begin{tabular}{l r r }
\toprule
\textbf{Epoch} & \multicolumn{1}{c}{\textbf{Train loss}} & \multicolumn{1}{c}{\textbf{Validation loss}} \\ 
\midrule
1              & 1.66                                     & 1.55                                          \\ 
2              & 1.47                                     & 1.34                                          \\ 
3              & 1.29                                     & 1.22                                          \\ 
4              & 1.18                                     & 1.16                                          \\ 
5              & 1.08                                     & 1.12                                          \\ 
6              & 1.00                                     & 1.09                                          \\ 
7              & 0.93                                     & 1.07                                          \\ 
8              & 0.87                                     & 1.06                                          \\ 
9              & 0.78                                     & 1.05                                          \\ 
10             & 0.72                                     & 1.06                                          \\ 
11             & 0.67                                     & 1.06                                          \\ 
12             & 0.62                                     & 1.07                                          \\ 
\bottomrule
\end{tabular}
\caption{Loss values during the Retrieval Augmented Fine-Tuning on the Human Dataset.}
\label{tab:human_loss}
\end{table}

\begin{figure}[hbt!]
\centering
\includegraphics[width=\linewidth]{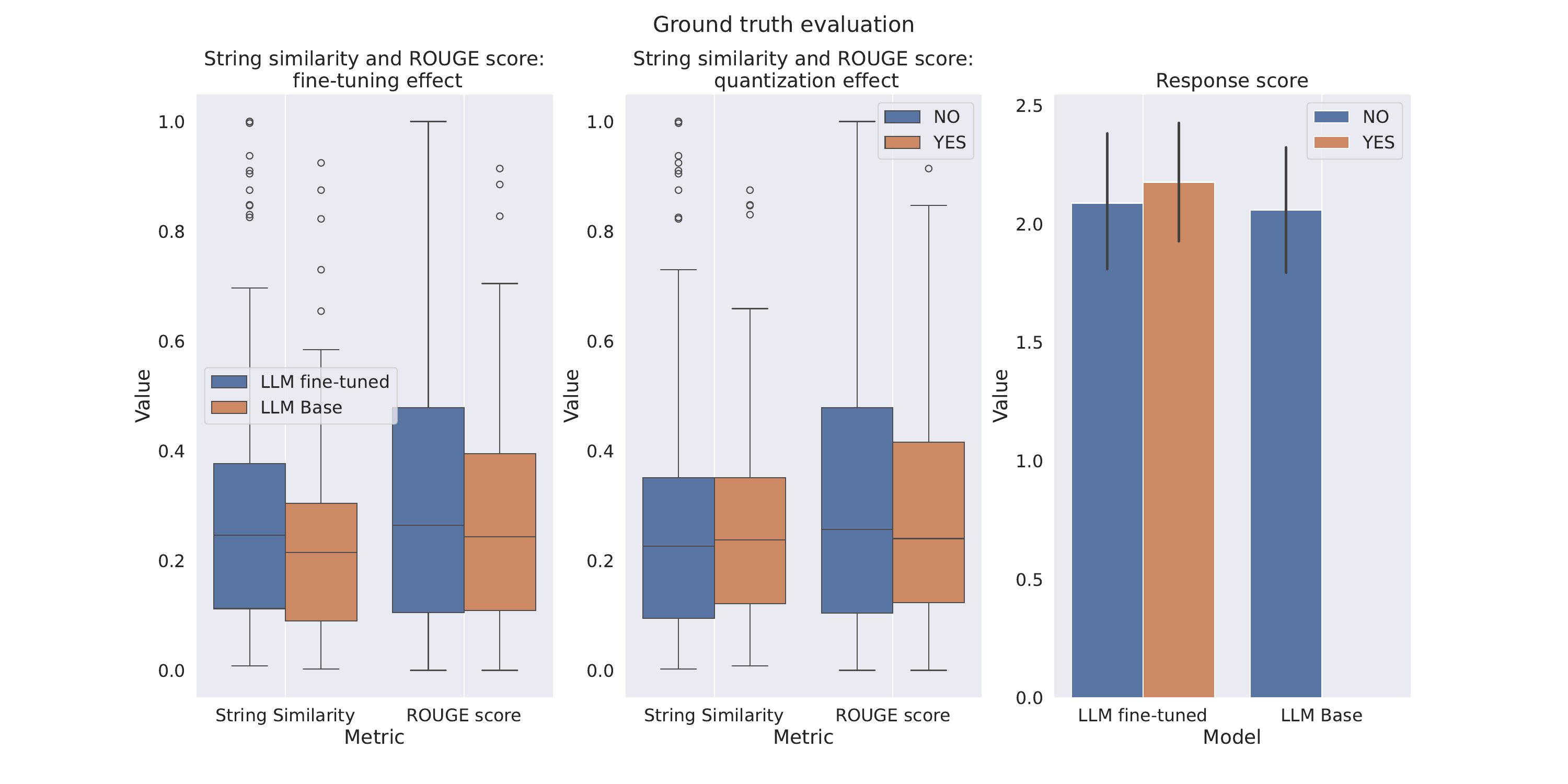}
\caption{Evaluation of the effect of fine-tuning and the quantization process on the LLM performance.}\label{fig:groundtruth_comparison_EvalDataset}
\end{figure}

The evaluation of the fine-tuning and the quantization procedure on the LLM performance showed mixed results  (Figure \ref{fig:groundtruth_comparison_EvalDataset}). The fine-tuned LLM yielded better results than its non fine-tuned version when considering both quantitative and qualitative metrics. After the quantization process the LLM performances decreased in term of ROUGE score, while increased in term of string similarity and response score (i.e., the human evaluation). 

Due to the requirement of memory efficiency (see \ref{phase1}) and the better score obtained in the human evaluation, we decided to use the quantized version of the fine-tuned LLM in the framework and for the testing activities.

\subsection{Evaluation of the first release}
In the first round of testing activities, the framework was prompted with a total of 108 (68 from the Evaluation Dataset) questions. In all these interactions, the testers did not select a single UNISIG specification, thus, widening the search of the retriever to all the available documents. The resulting average rating was 1.25. In detail, the 44\% of the responses were rated with the lowest score (i.e., 1), 13\% of the responses were evaluated with a score of 2, 20\% received a score of 3 and the highest score (i.e., 4) was assigned to 23\% of responses (see Figure \ref{fig:round_comparison}). Besides the scores, the Testing Group reported a drop in the quality of the responses after a few interactions (e.g., after three consecutive question-answer) and the occurrence of the default message (i.e., the implemented guardrail) even when it was not necessary. More in general, an increased occurrence of hallucinations has been observed when the question was about the meaning of an acronym and the results of the retrieval mechanism was poor. Few anomalies were also reported for the graphical interface, such as the wrong formatting of tables and long texts when displaying the chunks that generated the responses.

To address the issues reported by the Testing Group in this first round, we modified the embedding model, the retriever and the system prompt of the LLM (details in Appendix~\ref{app1}).  

\subsection{Evaluation of the second release}
The results of the second evaluation round showed a general performance improvement compared to the first round, with the average rating score increasing from 1.25 to 2.69. 
This improvement seems to be mostly driven by an increase (from 23\% to 39\%) of the responses rated with the highest score (i.e., 4) and a decrease  (from 44\% to 30\%) of the responses with the lowest score (i.e., 1).
On the other hand, no substantial changes were observed in the percentages of the responses with ratings 2 and 3 (see Figure \ref{fig:round_comparison}). 

The Testing Group noted a clear improvement in the retrieval mechanism, a better robustness of the framework in the management of medium and long conversations, and in the visualization of the default message (i.e., the guardrail). However, Figure \ref{fig:round_comparison} highlights that the framework seems to be biased towards the lower and upper boundary of the rating scale (i.e., 1 and 4), thus showing that the LLM can provide mainly  "exceptionally bad"  or "exceptionally good" responses. This result can be explained by the necessity, reported by the Testing Group, of further improvement in the retrieval mechanism. For instance, the framework provided satisfying responses for questions related to short and medium-sized tables, while reporting unsatisfying performance when dealing with long ones. Similarly, good performances were observed when the user question was related to contexts that are close within the technical standard, while additional work is required when the relevant contexts are sparse or their link is more subtle. Finally, the framework showed to be more prone to hallucinations when the retrieved contexts are partially or totally unrelated to the question, and, in these situations, simply rephrasing the question or providing clearer input did not always show to be effective.
No more anomalies were reported for the graphical interface.

To address the issues reported by the Testing Group in this second round, we modified the ingestion pipeline and the LLM (details in Appendix~\ref{app2}).

\subsection{Evaluation of the third release}
\begin{figure}[t]
\centering
\includegraphics[width=\linewidth]{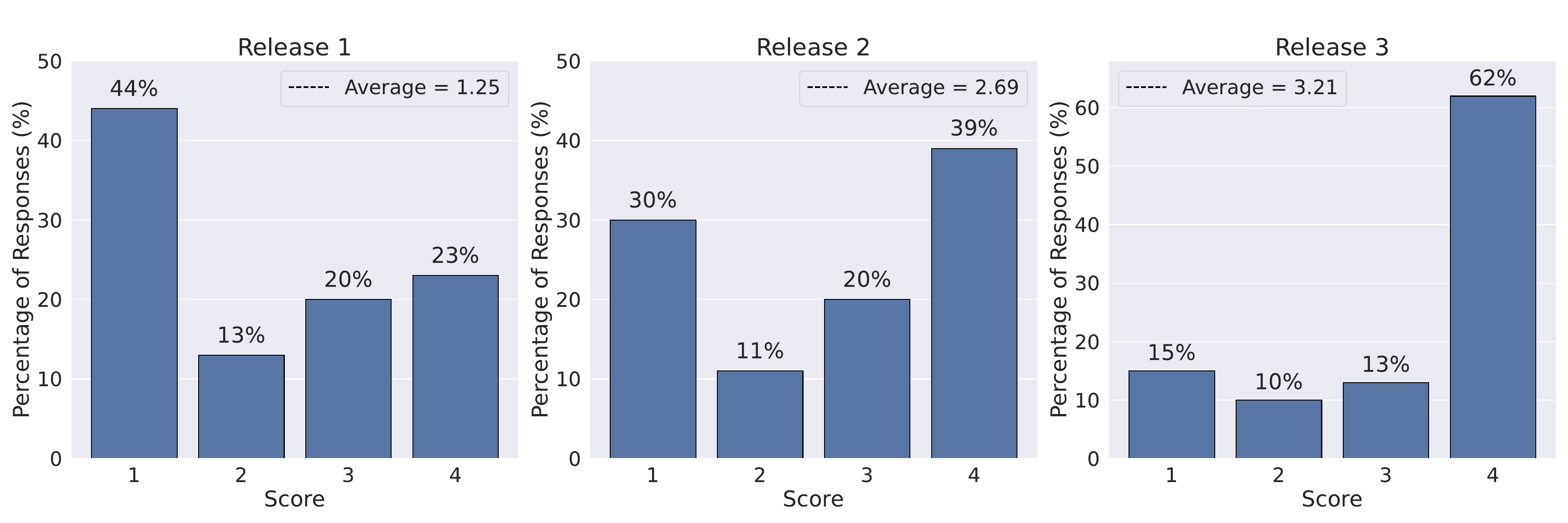}
\caption{Result comparison of the three framework releases.}\label{fig:round_comparison}
\end{figure}

The results of the third evaluation round showed a performance improvement compared to the previous rounds, with the average rating score increasing from 2.69 to 3.21. Major changes in the percentage were observed for the responses rated with 4 points (from 39\% to 62\%), the response rated with 1 point (30\% to 15\%) and the responses rated with 3 points (from 20\% to 13\%) (see Figure \ref{fig:round_comparison}). 

The Testing Group observed that the framework was more robust when dealing with retrieved context that was only partially relevant to the questions. Nevertheless, testers reported that the percentage of the unacceptable responses (i.e., responses rated with a score of 1) remains higher than expected, and in certain cases, the framework is still unable to provide satisfactory responses (see \ref{fig:round_comparison}). This result seems to be primarily related to the retriever's inability to effectively handle queries whose answers are not explicitly stated into the technical standards but require inference from multiple sparse information. 
Finally, the testers highlighted the need for further improvements in the handling of very long tables and emphasized the incorporation of methods to enable the interrogation of diagrams and images.

\section{Discussion}

In this work, the implementation of a framework based on a Retrieval-Augmented Generation approach to facilitate the consultation of technical regulations in the railway domain has been presented. 
While this implementation specifically targets UNISIG specifications, the framework has been deliberately designed with transferability as a core principle, making it readily adaptable to various technical domains.

The transferability of our solution is anchored in three key components. First, the document parsing pipeline can handle documents with complex layout and structure, including text, tables, and diagrams, that are common across technical documentation. Second, the embedding and retrieval mechanisms are domain-agnostic by design, requiring only new datasets on different contexts. Third, a reproducible fine-tuning approach using both synthetic and human-annotated datasets.


An important characteristic of our methodology was the continuous involvement of human experts throughout all phases. The creation of various specialized groups composed of industry professionals provided critical guidance and feedback that constantly shaped the framework's development, while the collaborative approach between domain expert engineers and academic researchers in AI represented an added value in the selection of the best approaches and tools in the current literature.

The framework features local deployment capabilities that are privacy-focused, regulation-compliant, and on-premise, and the implementation of countermeasures against privacy attacks. 
The modular design allows independent component upgrades and LLM quantization reduces computational requirements while maintaining performance.

\section{Limitations}
A critical limitation of this work is the lack of a comparative baseline. However, although we recognize the importance of a comparison with existing frameworks and solutions, the regulatory and privacy requirements of our use case prevented direct evaluation against commercial or external systems. Moreover, commercial tools typically rely on on external cloud infrastructures or API-based services, which are not compatible with our deployment context. 
The results in Figure~\ref{fig:round_comparison} show that the presented framework has acceptable performance, although there is room for improvement and to overcome some limitations. 
For instance, the Testing Group reported that the LLM can provide mainly  "exceptionally bad"  or "exceptionally good" responses, as showed in Figure~\ref{fig:round_comparison}. This result can be explained by the necessity of further improvement in the retrieval mechanism to deal with questions whose response is fragmented across the document or is contained in long tables. Moreover, the tester reported that the framework showed to be more prone to hallucinations when the retrieved contexts are partially or completely unrelated to the question, and, in these situations, simply rephrasing the question or providing clearer input did not always show to be not always effective.

Another important limitation of the framework is the impossibility of answering questions on images and diagrams contained in the document. However, due to the modular nature of the framework and the advancement in the AI field, this limitation can be addressed by introducing multi-modals models, both across the implementation pipeline and in the inference phase,  more tailored embedding models (see \cite{khanna2024tabularembeddingmodeltem} for an example) or the retriever mechanism (see for example \cite{ye2025infiniteretrievalattentionenhanced}). 

Another critical area for improvement is the management of the chat history, as it represents a constant increase in the memory loading that, especially in long conversations, could significantly degrade the performance \cite{sumida2024ragchatbotsforgetunimportant}. Similarly, the on-premise deployment could represent a critical issue when scaling the approach to much larger document collections. 
Finally, although the user-friendly rating system embedded in the graphical user interface provides an easy way to acquire data for both diagnostic purposes and future improvements, the framework would benefit from a more structured end-to-end quantitative evaluation of latency, helpfulness, and consistency of the generated response \cite{zeng2024knowledgecheckingretrievalaugmentedgeneration}.

 

\section{Conclusion}

From the development of this framework, several lessons emerged. First, early and continuous involvement of end-users proved to be essential in aligning technical choices with operational needs. Second, modularity and transparency in design facilitated iterative improvements and ensured compliance with strict privacy requirements. Finally, while generative AI offers promising opportunities for consulting technical standards, its deployment in safety-critical domains demands rigorous validation, domain-specific adaptation, and clear fallback strategies to mitigate risks.

Future work will focus on enhancing retrieval for multi-context reasoning, integrating multi-modal capabilities for diagrams and images, and exploring hybrid approaches that combine symbolic reasoning (rule-based or knowledge graphs) with generative models. By sharing this experience, we aim to provide practicable insights for small and medium enterprises seeking to leverage generative AI within regulated environments, balancing innovation with reliability and safety.

\section*{Declaration of generative AI and AI-assisted technologies in the manuscript preparation process}
The author(s) did not use any AI-assisted technology in producing the manuscript.
The author(s) only relied on grammar checking tools and take(s) full responsibility for the content of the published manuscript.

\section{Acknowledgment}
This work is part of the project "Generative Models: Empowering Business Processes and Enhancing Workflows for Improved Performance" (CUP: C49H23000090009) funded by the European Union - Next Generation EU.
The views and opinions expressed are those of the authors only and do not necessarily reflect those of the European Union or the European Commission. Neither the European Union nor the European Commission can be held responsible for them.

\bibliographystyle{plain} 
\bibliography{mybibfile.bib}

\appendix
\clearpage

\section{Modifications introduced in the second release}
\label{app1}
The issues reported by the Testing Group in first evaluation round, were addressed by modifying the embedding model, the retriever and the system prompt of the LLM.
The annotated chunks of the Human Dataset were used to evaluate the performance of the new retriever system with and without a new embedding model by measuring the recall, the precision, the string similarity, and the ROUGE score between each retrieved chunk and the available expected chunk. The precision is calculated as the proportion of relevant chunks in the retrieved contexts, while the recall indicates how many relevant chunks were retrieved. Therefore, given that there is only one relevant chunk (i.e., the chunk of the UNISIG specifications associated to a question-answer pair) and three retrieved chunks, the precision can range from 0 to 1, whereas the recall can be only 0 or 1 thus indicating if the relevant chunk is among the retrieved or not. Moreover, string similarity and ROUGE score vary in range between 0 and 1. However, since there is only one relevant context but three contexts are usually retrieved by the framework, only the maximum values reached by the string similarity and ROUGE score were considered in the analysis.
The default retriever system (i.e., pure semantic search) and default embedding model\footnote{\texttt{multi-qa-mpnet-base-dot-v1}}, which is an optimized version for search-style applications of the \texttt{mpnet-base-v1} initially introduced in \cite{song2020mpnetmaskedpermutedpretraining}) were compared with the newer version of the same embedding model\footnote{\texttt{mpnet-base-v2}} and with an ensemble retriever that combined semantic search with a term-based search, namely BM25 \cite{BM25}. 

\begin{figure}[!tb]
\centering
\includegraphics[width=\linewidth]{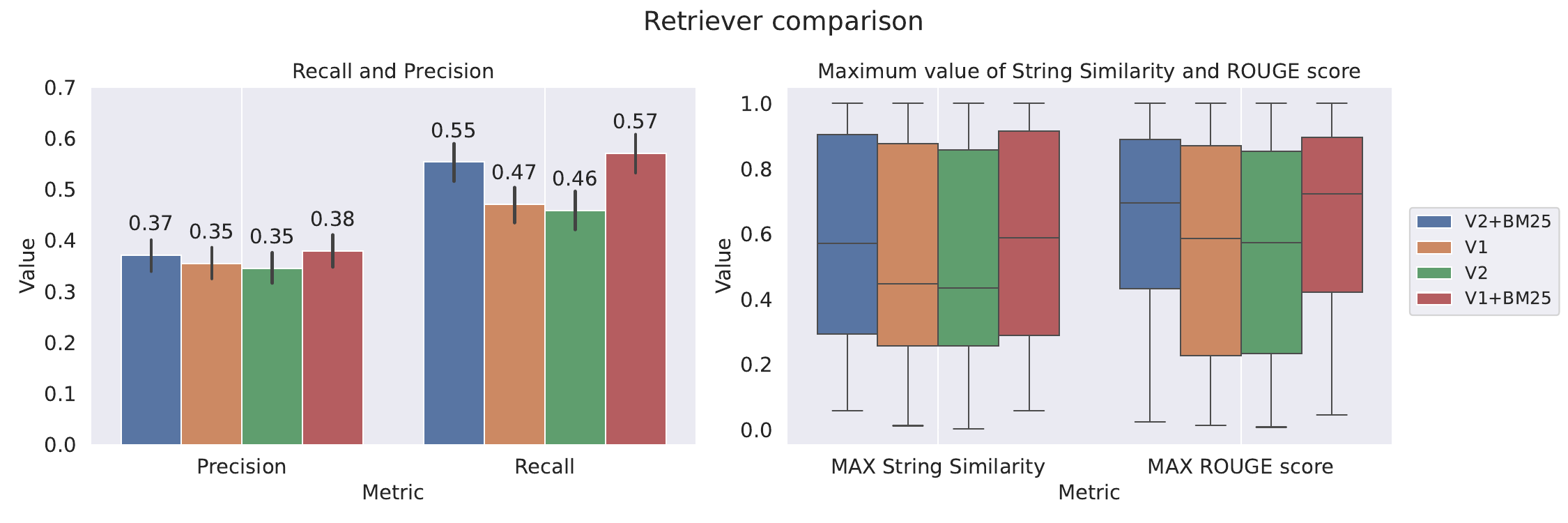}
\caption{Comparison of the both embedding model and retriever mechanism.}\label{fig:retriever_comparison}
\end{figure}

The analysis of the results showed that for all the metrics considered, the ensemble retriever outperforms the simpler semantic-based retriever independently of the embedding model used. On the other hand, the default embedding model yielded slightly better results than the newer version, although it is an older model (see Figure \ref{fig:retriever_comparison}).

Therefore, in the second version of the framework, an ensemble retriever coupled with the default embedding model has been used.

Finally, a chain-of-thoughts (CoT) approach, which has been observed to improve the LLM reasoning abilities~\cite{wei2023chainofthoughtpromptingelicitsreasoning}, was introduced in the system prompt to improve the quality of the responses and avoid the degradation observed after a few interactions (details in Appendices~\ref{app:cot} and~\ref{app:response_time}).

\section{Modifications introduced in the third release}
\label{app2}

The issues reported by the Testing Group in second evaluation round, were addressed by modifying the data ingestion process and by using a new LLM.
The data ingestion pipeline has been redesigned to enforce structural separation between tabular and textual content. Specifically, tables were processed as discrete chunks, each of them paired with the table caption and a brief manual-written overview. By isolating tables from textual information, we expected a mitigation of the noise during similarity scoring, ultimately improving the accuracy of retrieved content.
Leveraging the system modularity the LLM was changed. In detail, \textit{Zephyr7B-beta} has been replace by \textit{gemma-3-4b-it}, a lightweight instruction-tuned model designed to deliver support for multi-modal input, improved reasoning and instruction following capabilities compared to its predecessor \cite{gemmateam2025gemma3technicalreport}.

\section{Text extraction and post-processing pipeline}
\label{app:extraction}

\subsection{PDF parsing and content segmentation}
To extract structured content from PDF documents, we utilized the \textit{Unstructured} library with the YOLOX object detection model.
The configuration was defined as \lstinline{model="yolox"} and \lstinline{strategy="hi_res"} to enhance the resolution of page rendering.
Table structure inference was enabled via the \lstinline{infer_table_structure=True} parameter.

Each page of a PDF is converted into an image, which is then processed by the YOLOX model to detect and classify layout components such as titles, tables, images, and narrative text blocks.
The detection is performed by the \lstinline{partition_pdf} function, which returns a Python dictionary serializable as a JSON object.
Each element includes either a \lstinline{text} or \lstinline{text_as_html} attribute (the latter being present when table inference is enabled) and is labeled with a specific type, such as ``Title'', ``Table'', or ``NarrativeText''.

These elements are further processed using a custom script that exports them to plain text (.txt) files.
During this phase, a list of uninformative strings is filtered out, and page numbers are removed to clean the content.
Text is then segmented using the \lstinline{CharacterTextSplitter} class with the following parameters: \lstinline{chunk_size = 1000}, \lstinline{chunk_overlap = 200}, and \lstinline{separator = "\n\n"}.

The resulting chunks are loaded using the \lstinline{TextLoader} class from the \lstinline{document_loaders} module of \textit{LangChain}.
Each chunk is enriched with metadata from the corresponding JSON output, including attributes such as page number, paragraph identifiers, normative references, and content-specific IDs.

\subsection{Applications of enriched chunks}

\subsubsection{Vector store generation for RAG}
Each text chunk is embedded using the \texttt{multi-qa-mpnet-base-dot-v1} sentence embedding model.
The resulting vectors are stored in two types of vector databases to enable semantic retrieval:
\begin{itemize}
    \item A document-specific vector store for each input PDF.
    \item A global vector store aggregating all documents.
\end{itemize}
During evaluation, both FAISS and Chroma were considered as backend solutions for vector storage.
FAISS was ultimately selected for the final implementation due to its superior performance in retrieval speed and accuracy during internal benchmarking.

\subsubsection{CSV generation for web-based annotation}
In parallel with vector generation, each enriched chunk, along with its metadata, is exported into a CSV file format.
This file serves as input to a web-based annotation interface, enabling human reviewers to inspect and manually label the extracted content.
The inclusion of structured metadata facilitates efficient navigation and contextual understanding during the annotation process.

\subsection{Web-based annotation tool for dataset creation}
\label{app:annotation_tool}
A custom web-based annotation tool was developed to support the creation of a training dataset for question-answering tasks.
The application is built using PHP and the Laravel framework, which offers modern development practices, high code maintainability, and robust security features suitable for scalable web applications.

The tool presents enriched text chunks, previously extracted and segmented from PDF documents, to users, who are prompted to generate corresponding question-answer pairs.
Each user-generated pair is stored in a MySQL relational database, which serves as the back-end for managing annotation sessions and storing structured data.
This database also retains metadata related to the source chunk, enabling traceability and contextual linkage for downstream model training and evaluation.

\section{Datasets for fine-tuning}
\label{app:datasets}

\subsection{Augmented Dataset}
\begin{figure}[!htb]
\centering
\includegraphics[width=\linewidth]{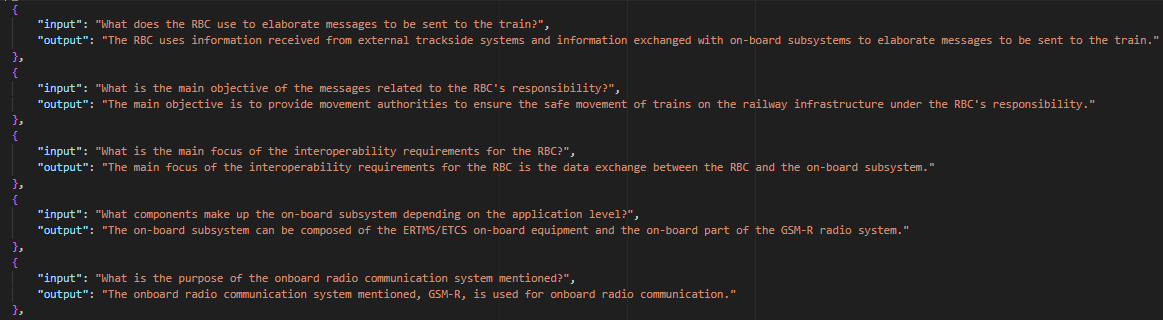}
\caption{Examples extracted from the augmented dataset.}\label{fig:supp_augmented_examples}
\end{figure}

The augmented dataset was generated by leveraging the Qwen14B model, which was used to create question-answer pairs from extracted phrases taken from the normative texts.
The model was prompted to ``create a question-answer pair based on this context'' for each phrase.
This process automatically generated question-answer pairs that were then formatted for training (Figure~\ref{fig:supp_augmented_examples}).

Input format:
\begin{lstlisting}
Inputs = "Question: {dataset_question} \n"
Targets = "{answer} <\s>"
\end{lstlisting}
The end-of-sequence token \lstinline{<\s>} was appended to the target to help the model learn where to stop generating tokens.
Tokenization was performed using the following method:
\begin{lstlisting}
tokenized = tokenizer(inputs, text_target=targets, max_length=512,
                      truncation=True, padding="max_length")
\end{lstlisting}

Alternative preprocessing strategies were also explored.
For instance, we experimented with a conversational template format using role-based tags (Figure~\ref{fig:supp_augmented_alternative}).
Alternative input format:
\begin{lstlisting}
<user> {Question} \n
<assistant> {answer}
\end{lstlisting}

\begin{figure}[!htb]
\centering
\includegraphics[width=\linewidth]{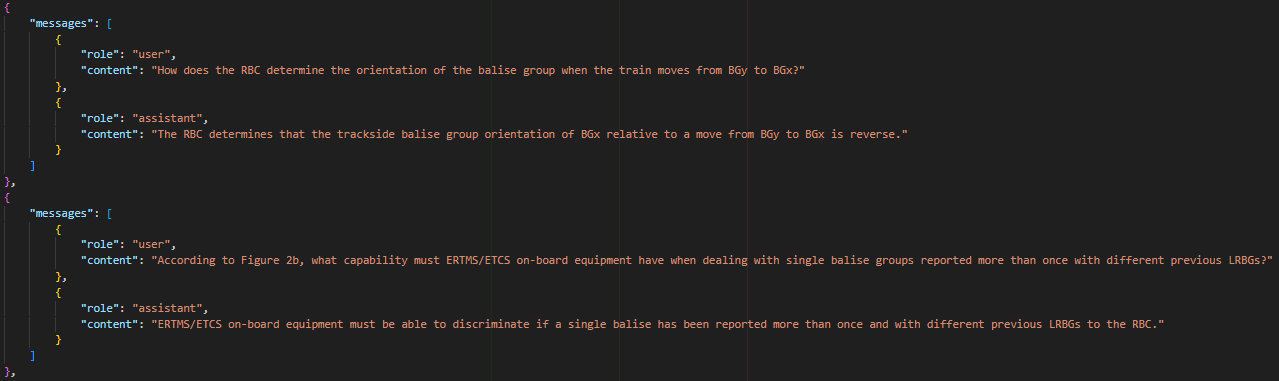}
\caption{Examples of the alternative augmented dataset.}\label{fig:supp_augmented_alternative}
\end{figure}

However, empirical evaluation showed that this format led to reduced performance compared to the standard prompt structure.

\subsection{Human Dataset}
In the second fine-tuning stage, a human-labeled dataset was used to teach the model how to leverage context in producing accurate answers.
The structure was similar to the augmented dataset but included an explicit context field.
This dataset was collected via the web-based annotation tool (Appendix~\ref{app:annotation_tool}), where the context was shown to users, and they were prompted to generate a corresponding question and answer.

Input format with context:
\begin{lstlisting}
Inputs = "Context: {context} \nQuestion: {dataset_question} \n"
Targets = "{answer} <\s>"
\end{lstlisting}
Tokenization followed the same approach as in the augmented dataset, with padding and truncation applied:
\begin{lstlisting}
tokenized = tokenizer(inputs, text_target=targets, max_length=512,
                      truncation=True, padding="max_length")
\end{lstlisting}

Example from the human dataset:
\begin{lstlisting}
Context:
"If sufficient resources are available to provide the service (in both the local and distant system) the new connection will be established.

Otherwise the connection request is refused.

8.2.4.2.5 The handling of transport priority during the data phase of the transport connection is specified in the following section.

8.2.4.3 Multiplexing

8.2.4.3.1 Multiplexing of two or more transport connections onto a single network connection can be provided as an option. This option is not required for ETCS level 1 radio in-fill unit.

8.2.4.3.2 Multiplexing requires the following functions:

a) The identification of the transport connection source is provided by an appropriate DST-REF parameter of each DT TPDU and additionally the SRC-REF parameter of CR, CC, DR, and DC TPDUs. These parameters are used to identify each TPDU in a given transport connection and ensures that data from different transport connections are not mixed or mis-routed.

', '3.13.2.2.9.1.3 For a given confidence level on emergency brake safe deceleration (EBCL), the rolling stock correction factor Kdry_rst(V) shall be given as a step function of speed, with the same steps as the ones of A_brake_emergency(V).

3.13.2.2.9.1.4 The confidence level on emergency brake safe deceleration represents the probability of the following individual event: the rolling stock emergency brake subsystem of to A_brake_emergency(V) * Kdry_rst(V), when the emergency brake is commanded on dry rails.

3.13.2.2.9.1.5 The rolling stock correction factor Kwet_rst(V) shall be given as a step function of speed, with the same steps as the ones of A_brake_emergency(V). It represents the loss of deceleration with regards to emergency braking on dry rails, when the emergency brake is commanded on wet rails, according to wheel/rail adhesion reference conditions.

3.13.2.2.9.2 Correction factor for gradient on normal service deceleration"

Question: How does the onboard equipment communicate the Override?

Answer: In level2 and 3, the onboard equipment communicate the Override has been triggered by means of the reported mode change (if there is any and if radio communication is available). Also, if the ERTMS/ETCS-onboard equipment is able to report the mode change to the RBC, the RBC may transmit limits for the distance to run in SR mode (overriding the national value), a list of balises to be passed in SR mode.
\end{lstlisting}

\section{Training parameters}
\label{app:training}
This work presents a two-stage fine-tuning approach for progressively adapting a large language model to a specialized domain.
In the first stage, the model is fine-tuned on an Augmented Dataset containing a mix of synthetic and curated examples to establish a broad domain foundation (Table~\ref{tab:supp_params_augmented}).
The second stage further refines the model using Retrieval-Augmented Fine-Tuning (RAFT) on a high-quality Human Dataset (Table~\ref{tab:supp_params_human}), enhancing its precision and contextual understanding.

Both stages use LoRA for parameter-efficient training but differ in key hyperparameters such as batch size, learning rate, and gradient accumulation, reflecting the distinct nature of the datasets.
This pipeline demonstrates how combining synthetic pre-specialization with human refinement can lead to more robust domain adaptation.

\begin{table}[!htb]
\centering
\begin{tabular}{l r}
\toprule
\textbf{Parameter} & \textbf{Value} \\
\midrule
Number of epochs & 20 \\
Batch size & 4 \\
Gradient accumulation & 16 \\
Initial learning rate & 2E-4 \\
Warmup steps & 2 epochs \\
Learning rate decay & 0.01 \\
Early stopping & Yes \\
Learning rate scheduler & Linear \\
\midrule
\multicolumn{2}{c}{\textbf{LoRA parameters}} \\
\midrule
Rank & 32 \\
Dropout & 0.1 \\
Alpha & 64 \\
Trained layers & All \\
\bottomrule
\end{tabular}
\caption{List of parameters used in the supervised fine-tuning on the Augmented Dataset.}
\label{tab:supp_params_augmented}
\end{table}

\begin{table}[!htb]
\centering
\begin{tabular}{l r}
\toprule
\textbf{Parameter} & \textbf{Value} \\
\midrule
Number of epochs & 20 \\
Batch size & 16 \\
Gradient accumulation & No \\
Initial learning rate & 1E-5 \\
Warmup steps & 2 epochs \\
Learning rate decay & 0.01 \\
Early stopping & Yes \\
Learning rate scheduler & Linear \\
\midrule
\multicolumn{2}{c}{\textbf{LoRA parameters}} \\
\midrule
Rank & 32 \\
Dropout & 0.1 \\
Alpha & 64 \\
Trained layers & All \\
\bottomrule
\end{tabular}
\caption{List of parameters used in the Retrieval-Augmented Fine-Tuning on the Human Dataset.}
\label{tab:supp_params_human}
\end{table}

\section{Model quantization process}
\label{app:quantization}
The model was quantized using the \textit{BitsAndBytes} Python library, which offers efficient quantization and GPU optimization capabilities, particularly suitable for large language models.
By combining quantization with PEFT, we performed QLoRA training.
This approach requires the \textit{accelerate} Python library to ensure efficient GPU utilization during both training and inference.

BitsAndBytes configuration:
\begin{lstlisting}
bnb_config = BitsAndBytesConfig(
    load_in_4bit=True,                       # 4-bit quantization
    bnb_4bit_quant_type='nf4',               # Normalized float 4
    bnb_4bit_use_double_quant=True,          # Second quantization
    bnb_4bit_compute_dtype=torch.bfloat16    # Computation type
)
\end{lstlisting}
The computation type was set to \lstinline{torch.bfloat16} because the L40 Nvidia GPU used supports this numeric format, which is more efficient than standard float16.

\paragraph{Limitations of quantization.}
\begin{itemize}
    \item Even advanced GPUs cannot perform arithmetic operations in native 4-bit format.
    \item Matrix multiplication and other operations require higher precision to prevent catastrophic accuracy loss.
    \item 4-bit precision is too coarse for reliable gradient flow and numerical stability during training or inference.
    \item Inference speed may be slower due to conversion operations (from 4-bit to 16-bit for computation and back).
\end{itemize}

PEFT configuration:
\begin{lstlisting}
peft_config = LoraConfig(
    lora_alpha=64,             # rank * 2
    lora_dropout=0.1,
    target_modules="all-linear",
    r=32,
    bias="none",
    task_type="CAUSAL_LM",
)
\end{lstlisting}

\section{Guardrail implementation}
\label{app:guardrail}
The guardrail mechanism has been implemented to ensure that the language model responds strictly within the railway domain, preventing generic, off-topic, or unauthorized content.
The guardrail check is performed during the RAG chain pipeline, both in the retrieval phase and in the answer generation phase.

The first check, performed during the retrieval phase, consists of computing embeddings for the query and the relevant text chunks obtained through the retriever.
Using cosine similarity, the system evaluates how well the query aligns semantically with each retrieved text chunk, then calculates the average cosine similarity.
If the average similarity score falls below a defined threshold, the system raises a \lstinline{GuardrailError}, rejecting the query.
This allows the system to proactively identify incorrect, vague, or inappropriate questions that may not align with the available database.
In cases of otherwise coherent questions, it also helps detect issues strictly related to the retriever, such as irrelevant or poorly matched text chunks.

The second check occurs after the answer generation phase and is performed between the generated response and the retrieved text chunks.
By computing embeddings for both the response and the text chunks and evaluating the cosine similarity, the system assesses how well the response aligns semantically with the relevant text.
If the average similarity score falls below the defined threshold, the system raises a \lstinline{GuardrailError}, rejecting the response.
This helps prevent issues such as hallucinations, off-topic answers, or responses based on incorrect reasoning.

\lstinline{GuardrailError} is a custom exception class that inherits from Python's built-in \lstinline{Exception} and includes a user-friendly message shown to the user when a guardrail check fails.

The selected thresholds, determined after a testing phase, are:
\begin{itemize}
    \item 0.3 for the retrieval phase guardrail;
    \item 0.5 for the post-generation guardrail.
\end{itemize}
These values were chosen to be strict enough to exclude the majority of undesirable behaviors originating from the model, the retriever, or the user.

\section{Graphical interface and user experience}
\label{app:gui}
The tool is a web-based application, and the front-end has been developed using PHP, JavaScript, and CSS with the Laravel-Filament framework.
This widely used framework ensures high code standardization and follows the most up-to-date best practices in terms of cybersecurity and scalability.
The tool is linked to a Python-developed backend, where the LLM deployment, LLM-related queries management, and the retriever databases are located.
Through this application, users can instantiate a chat session with the model and choose the main topic from a list of documents composing the UNISIG database.
After receiving an answer from the model, users can consult the text chunk analyzed by the model and optionally download the related file used to provide the answer.
Additionally, users can vote and comment on the answer, which helps developers gather feedback to ensure continual improvements.

The application's database was implemented using the MySQL framework.
This database contains all the necessary information for correct deployment and future testing purposes.
For each chat session created by the users (previously anonymized), the database stores the title, the question-answer pairs, the related text chunks for each pair, as well as votes and comments.

\subsection{Workflow and user interaction}
\begin{figure}[!htb]
\centering
\includegraphics[width=\linewidth]{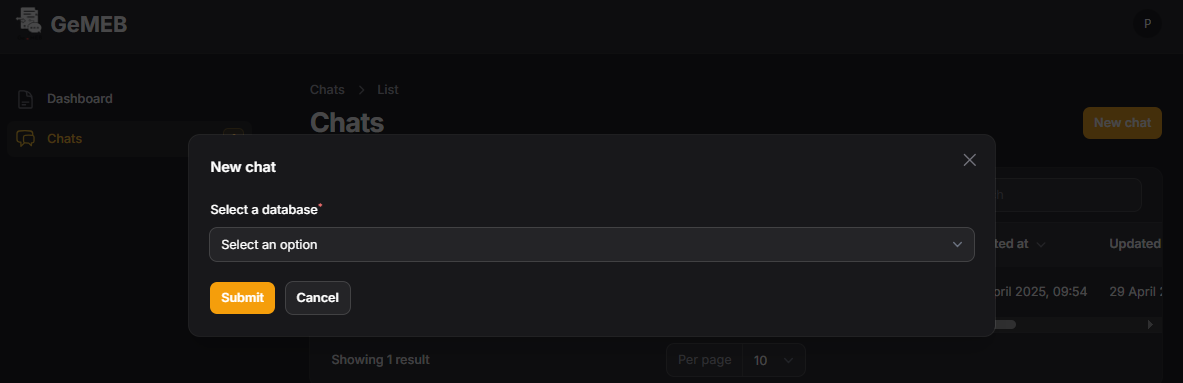}
\caption{Database selection interface in the application.}\label{fig:supp_gui_database_selection}
\end{figure}

Once a user has logged in, they can access the ``Chats'' page.
By clicking the ``New Chat'' button, they must select a database from a pre-defined list and initiate a chat session with the LLM (Figure~\ref{fig:supp_gui_database_selection}).

\begin{figure}[!htb]
\centering
\includegraphics[width=\linewidth]{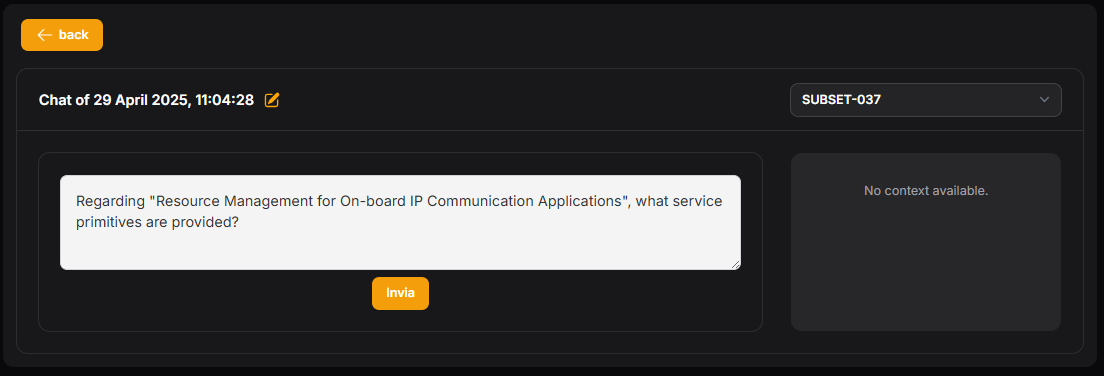}
\caption{Asking a question to the chatbot.}\label{fig:supp_gui_question}
\end{figure}

Once the database is selected, users can ask questions to the chatbot regarding the selected topic/database (Figure~\ref{fig:supp_gui_question}).

\begin{figure}[!htb]
\centering
\includegraphics[width=\linewidth]{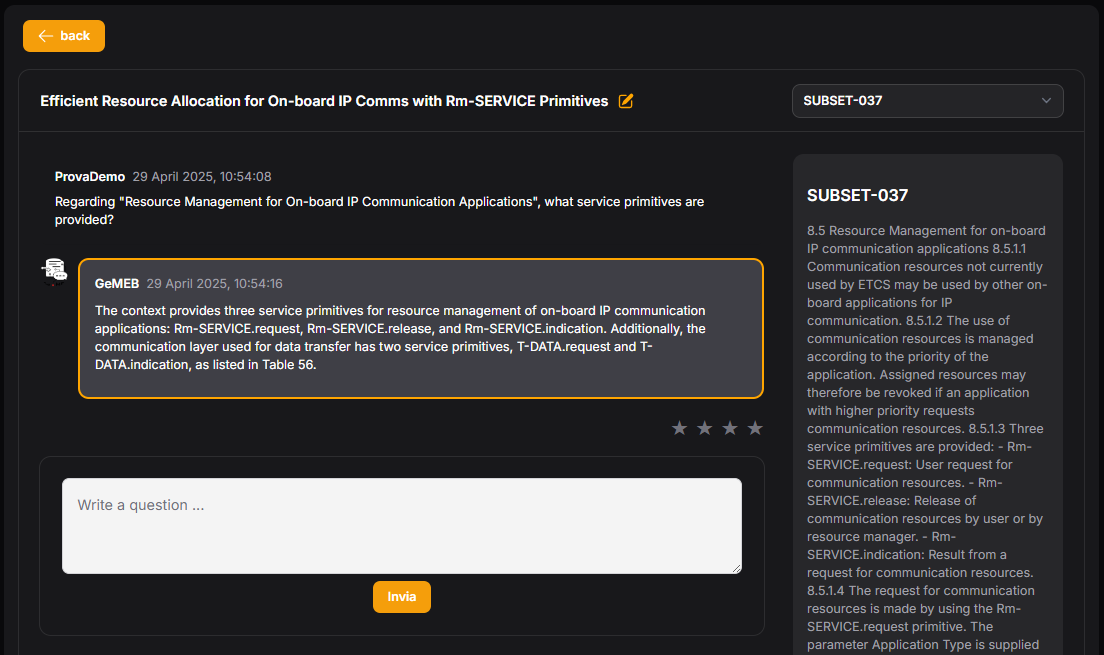}
\caption{LLM response and text chunk display.}\label{fig:supp_gui_response_chunks}
\end{figure}

After the question is sent, the Python-developed backend processes it.
If the guardrail checks pass (Appendix~\ref{app:guardrail}), the LLM answer is provided to the user.
The user can also view the question-answer timestamps and the text chunks identified by the retriever, which were used by the LLM to generate the answer (Figure~\ref{fig:supp_gui_response_chunks}).
During the first interaction, the LLM also generates a chat title, which is modifiable by the user.

\begin{figure}[!htb]
\centering
\includegraphics[width=0.75\linewidth]{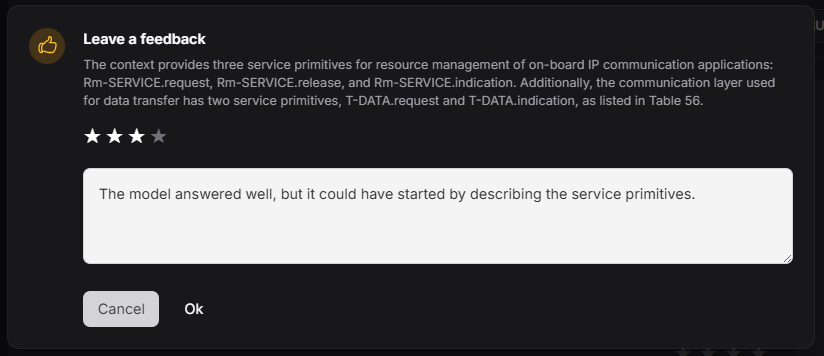}
\caption{Rating the LLM answer.}\label{fig:supp_gui_rating}
\end{figure}

The user can rate the answer provided by the LLM by clicking on the rating stars (Figure~\ref{fig:supp_gui_rating}).

\begin{figure}[!htb]
\centering
\includegraphics[width=0.7\linewidth]{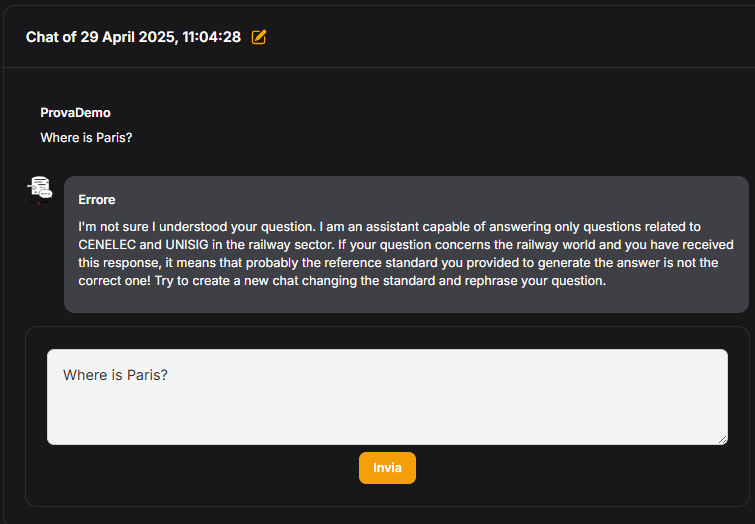}
\caption{Guardrail default answer.}\label{fig:supp_gui_guardrail}
\end{figure}

Subsequently, users can continue the conversation with the model.
If the user's question triggers the guardrail, the default answer shown in Figure~\ref{fig:supp_gui_guardrail} is provided in response.

\section{Chain of thoughts}
\label{app:cot}
The Chain of Thoughts (CoT) technique is widely adopted in the field of large language models (LLMs) because it encourages step-by-step reasoning, improves the interpretability of model decisions, and often leads to more accurate answers, especially for complex or multi-step queries.

In our implementation, it was necessary to adapt the prompt by removing the explicit use of chat history and instead relying on an LLM-generated reformulation that combines the user's question and the chat history.
The same reformulated query is used during the retrieval phase, enabling the system to generate responses that feel more natural and conversational by understanding the broader context of the interaction, rather than treating each question in isolation.

Due to the increased number of tokens required to generate the CoT, the maximum number of generated tokens was increased from 512 to 1024.

\paragraph{Prompt before CoT.}
\begin{lstlisting}
You are an AI specialized exclusively in the railway industry. Your role is to answer questions and provide information strictly related to the railway sector, including operations, technology, safety, and industry practices.

Guidelines:

1. Use the provided Context for detailed answers and History for context in follow-up questions.

2. Respond concisely and only answer questions related exclusively to railway topics.
   If the question is not related to the railway industry, respond with:
   "I can only provide information related to the railway industry."

3. Ensure all responses are concise, accurate, and focused exclusively on railway-related topics.

4. Maintain a professional, respectful tone. Avoid toxic or inappropriate language.

5. Do not generate or ask new questions.

Context: {context}

History: {history}

Question: {question}

Helpful Answer:
\end{lstlisting}

\paragraph{Prompt using CoT.}
\begin{lstlisting}
<|system|>You are an AI assistant specialized in the railway industry.

- Use the information contained in the Context to answer the Question.
- Think step-by-step before answering.
- Ensure all responses are concise and accurate.
- Format technical explanations using bullet points or markdown where appropriate.
- Do not include in the final answer: the question or parts of it, phrases like
  "In the given context", "The context explains", or similar.

<|user|>

Context:

{context}

Question:

{question}

First, analyze the question step-by-step by identifying relevant information in the context.

Then, provide your final answer on a new line starting with: "Final Answer:"

<|assistant|>

Thought Process:
\end{lstlisting}

\paragraph{Examples.}
\textbf{Question 1:} Provide me informations about Adaptation Layer Packet Format (no re-elaboration)

\textbf{Thought Process 1:}
\begin{enumerate}
    \item Identify the section in the context that provides information about Adaptation Layer Packet Format (ALEPKT).
    \item Understand the structure of ALEPKT and its components.
    \item Identify the significance of each component in the packet format.
    \item Understand the role of Adaptation Layer in handling packets in TCP/IP connections.
\end{enumerate}

\textbf{Answer 1:}
The Adaptation Layer Packet Format (ALEPKT) is used by the Adaptation Layer to delimit information exchanged by TS-Users in TCP/IP connections.
Each ALEPKT consists of a Packet Header and a User Data field.
The Packet Header contains the following components:
\begin{itemize}
    \item Packet Length: Indicates the length of the entire ALEPKT in octets, excluding the Packet Length field itself.
    \item Version: Used to identify the facilities offered by the Adaptation Layer to other parties. This parameter is agreed upon during configuration and may be ignored by the receiving ALE.
    \item Application Type: Identifies the application type, as specified in Subset-037 \S8.2.4.6. The Application Type of the calling and called peers shall be the same on both peers for a specific connection.
    \item T-Sequence number: Transport Sequence Number used by the receiving Adaptation Layer to manage switching between two active TCP connections. This field is not used for class A.
    \item N/R Flag: Used by the sending ALE to specify the normal or redundant links on which the ALEPKTs are sent.
    \item Packet Type: Describes the type of packet. An unrecognized Packet Type causes the ALE to discard the packet and possibly close the TCP connection.
    \item Checksum: Calculated using CRC-CCITT ($x^{16} + x^{12} + x^{5} + 1$) to detect data corruption.
\end{itemize}
The User Data field contains any legitimate length of user data.
The Adaptation Layer reassembles information packets from the byte stream at the TCP level to match the ISO Transport Service's TPDU boundaries.

\textbf{Question 2:} Can you provide me more informations about the Checksum field of ALEPKT?

\textbf{Re-elaboration:} How is the Checksum field calculated in the Adaptation Layer Packet Format (ALEPKT), and what is its purpose in ensuring the integrity of transmitted data?

\textbf{Thought Process 2:}
\begin{enumerate}
    \item Identify the relevant information in the context related to the Checksum field in the Adaptation Layer Packet Format (ALEPKT).
    \item Understand the calculation method for the Checksum field.
    \item Determine the purpose of the Checksum field in ensuring the integrity of transmitted data.
\end{enumerate}

\textbf{Answer 2:}
The Checksum field in the Adaptation Layer Packet Format (ALEPKT) is a 2-octet field used to calculate a cyclic redundancy check (CRC) value for the first 6 fields of the packet header, which includes the Packet Length, Version, Application type, T-Sequence number, N/R Flag, and Packet Type fields.
This CRC value is calculated using the generator polynomial specified in ISO/IEC 3309, which is CRC-CCITT ($x^{16} + x^{12} + x^{5} + 1$).
The purpose of the Checksum field is to ensure the integrity of transmitted data by identifying if any characters have been lost during transmission.
If the received Checksum value does not match the calculated value, the receiving Adaptation Layer may discard the packet and potentially close the TCP connection.
This helps to prevent errors and corruption in the transmitted data, ensuring its accuracy and reliability.

\textbf{Question 3:} What is the adaptation layer you mentioned before?

\textbf{Re-elaboration:} What is the Adaptation Layer and how does it help in managing the difference between TCP and ISO Transport Service in data transmission?

\textbf{Thought Process 3:}
\begin{enumerate}
    \item Understand the concept of Adaptation Layer and its role in managing the difference between TCP and ISO Transport Service in data transmission.
    \item Identify the relevant sections from the context that explain the Adaptation Layer and its functionality.
    \item Analyze the tables provided in the context to understand how the Adaptation Layer maps the ISO Transport Class 2 primitives to TCP/IP protocol.
\end{enumerate}

\textbf{Answer 3:}
The Adaptation Layer is a functional entity that bridges the gap between the Transport Service offered by the ISO Transport Service (ITSS) and the TCP/IP protocol used in data transmission.
It helps in managing the difference between these two protocols by providing a transport service interface that corresponds to that of ITSS but achieves the functionality of ISO Transport through the use of TCP/IP protocol.
The Adaptation Layer realizes the correspondence between the ITSS service and the primitives of TCP/IP protocol, maps the ITSS primitives to TCP/IP protocol, and manages the redundancy of TCP connections and the corresponding functional requirements.
This allows TS-Users to access the transport service provided by the Adaptation Layer Entity, which offers a Transport Class 2 service but achieves the functionality of ISO Transport through the use of TCP/IP protocol.
The Adaptation Layer helps in ensuring interoperability between end systems and preserving the behavior of the system, even when implementing modifications to adapt the ITSS user interface to implementation needs and constraints.

\section{Response time: base model vs CoT model}
\label{app:response_time}
We evaluated the response time of the model with and without the Chain of Thoughts (CoT) implementation (Appendix~\ref{app:cot}).
The average response time increased by approximately 2 seconds when CoT was enabled.
This increase is expected, as the reasoning process introduces additional steps that enhance response accuracy.

Despite the overall average increase, we observed that the 75th percentile of response times decreased with CoT.
This indicates that for the majority of queries, the model was able to structure its reasoning more effectively, resulting in faster answers.

\paragraph{Token analysis and correlation with response time.}
We also investigated the correlation between token counts and response time:
\begin{itemize}
    \item \textbf{Input tokens (prompt + retrieved context):} No significant correlation with response time was observed.
    \item \textbf{Output tokens:} Response time is strongly correlated with the number of generated output tokens.
\end{itemize}

\begin{figure}[!htb]
\centering
\includegraphics[width=\linewidth]{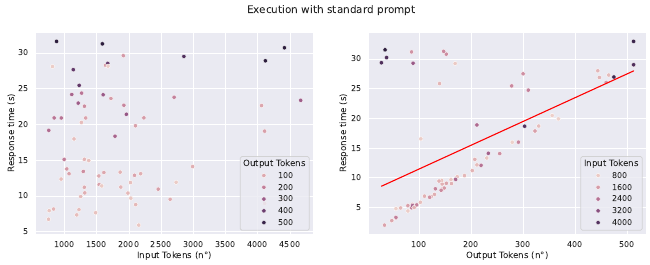}
\caption{Relationship between number of tokens and response time with standard prompt.}\label{fig:supp_tokens_standard}
\end{figure}

\begin{figure}[!htb]
\centering
\includegraphics[width=\linewidth]{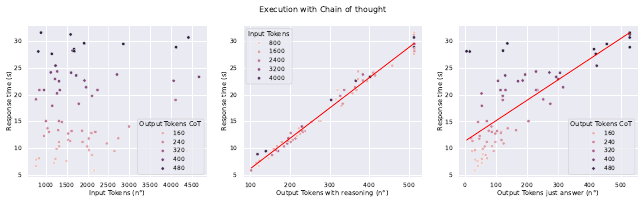}
\caption{Relationship between number of tokens and response time with CoT implementation.}\label{fig:supp_tokens_cot}
\end{figure}

For both models (with and without CoT), we observed a clear \textbf{linear relationship} between response time and the number of generated tokens (Figures~\ref{fig:supp_tokens_standard} and~\ref{fig:supp_tokens_cot}).
However, models without CoT showed several outliers, that is, instances where response time was disproportionately high even with a low token count.
These anomalies are likely due to poor comprehension of the input query or retrieved contexts, resulting in inefficient processing or irrelevant answers.

In contrast, the CoT-enabled model demonstrated an almost perfect linear correlation between response time and the total output length (including the reasoning steps and final answer).
This suggests that:
\begin{itemize}
    \item Reasoning steps in CoT models are more structured and predictable.
    \item Token generation becomes more consistent and aligned with processing complexity.
\end{itemize}

\paragraph{Impact of the thinking process.}
The rightmost plot in Figure~\ref{fig:supp_tokens_cot} shows that even when the generated answer is short, the total computation cost (in terms of response time) still scales with the CoT process.
This indicates that:
\begin{itemize}
    \item The reasoning component (even when not explicitly shown in the final answer) significantly contributes to the model's performance.
    \item The inherent overhead of step-by-step reasoning affects runtime, regardless of the length of the output answer.
\end{itemize}

\begin{figure}[!htb]
\centering
\includegraphics[width=\linewidth]{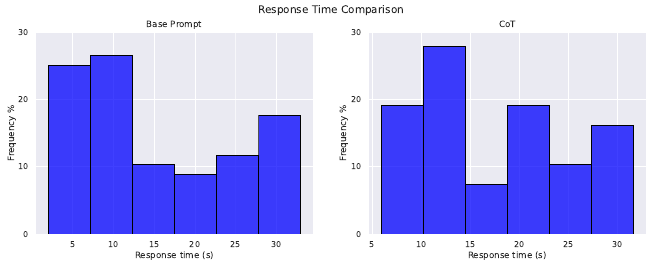}
\caption{Response time histograms comparison.}\label{fig:supp_response_time_hist}
\end{figure}

\begin{table}[!htb]
\centering
\begin{tabular}{l r r}
\toprule
\textbf{Statistic} & \textbf{Standard Prompt} & \textbf{CoT} \\
\midrule
count & 68 & 68 \\
mean & 15.34 s & 17.47 s \\
std & 9.69 s & 7.46 s \\
min & 2.03 s & 5.91 s \\
25\% & 7.13 s & 11.23 s \\
50\% & 12.13 s & 15.11 s \\
75\% & 25.90 s & 23.44 s \\
max & 32.96 s & 31.62 s \\
\bottomrule
\end{tabular}
\caption{Descriptive statistics for response times: Standard Prompt vs.\ CoT.}
\label{tab:supp_response_time_stats}
\end{table}

Statistical analysis was conducted to compare the response time distributions between the standard prompt and the Chain-of-Thought (CoT) implementation (Figure~\ref{fig:supp_response_time_hist} and Table~\ref{tab:supp_response_time_stats}).
Normality tests using the Shapiro--Wilk test indicated that neither distribution follows a normal distribution ($p < 0.05$ for both cases), which justifies the use of non-parametric testing methods.
The Mann--Whitney U test yielded a p-value of 0.053, which is slightly above the conventional significance threshold of 0.05.
Therefore, while there is a visible trend toward longer response times with CoT, the difference is not statistically significant at the 5\% level.
This suggests that, although CoT introduces additional processing steps, it does not significantly degrade performance in terms of response time.

\end{document}